\pdfoutput=1
\documentclass[%
reprint,
superscriptaddress,
nofootinbib,
amsmath,amssymb,
aps,
]{revtex4-1}

\usepackage{graphicx}

\usepackage{bm}
\usepackage{hyperref}

\usepackage[utf8]{inputenc}
\usepackage{graphicx}
\usepackage{bm}
\usepackage{hyperref}

\usepackage[utf8]{inputenc}

\usepackage{amsmath,amssymb}

\usepackage[T1]{fontenc} 
\usepackage{microtype} 
\usepackage[english]{babel} 
\usepackage{booktabs} 
\usepackage[utf8]{inputenc}
\usepackage{amsmath}
\usepackage{graphicx}
\usepackage{amssymb}
\usepackage{braket,mleftright}
\usepackage{empheq}
\usepackage{subfigure}
\usepackage{stackrel}
\usepackage{soul}
\usepackage{blkarray}
\usepackage{multirow}
\usepackage{amsmath}
\usepackage{physics}
\usepackage{amsfonts}
\usepackage{bm}
\usepackage{bbold} 
\usepackage{color}

\usepackage[utf8]{inputenc} 
\usepackage{amsmath}

\usepackage{enumitem} 
\setlist[itemize]{noitemsep} 

\usepackage{hyperref} 
\usepackage{braket}
\usepackage{verbatim} 

\begin{document}

\title{Navigating on Quantum Control Solution Subspaces}

\author{Mart\'{i}n Larocca}
\email{mail to: larocca@df.uba.ar}

\affiliation{Departamento de F\'{i}sica “J. J. Giambiagi” and IFIBA, FCEyN, Universidad de Buenos Aires, 1428 Buenos Aires, Argentina
}
\author{Esteban Calzetta}
\author{Diego Wisniacki}

\affiliation{Departamento de F\'{i}sica “J. J. Giambiagi” and IFIBA, FCEyN, Universidad de Buenos Aires, 1428 Buenos Aires, Argentina
}%

\date{November, 2019}%

\begin{abstract}

Quantum Optimal Control (QOC) is the field devoted to the production of external control protocols that actively guide quantum dynamics. Solutions to QOC problems were shown to constitute continuous submanifolds of control space. A solution navigation method exploiting this property to achieve secondary features in the control protocols was proposed [Larocca \textit{et al}, arXiv:1911.07105]. Originally, the technique involved the computation of the exact Hessian matrix. In this paper, we show that the navigation can be alternatively performed with a finite-difference approximation scheme, thus enabling the application of this procedure in systems where computing the exact Hessian is out of reach.

\end{abstract}

\maketitle

\section{Introduction} \label{Section-Intro}

The second quantum revolution is expected to deliver new technology harvesting fundamentally quantum properties like entanglement and superposition, with applications in simulation, computing, sensing and communication \cite{gis,bib:nori2014,bib:martinis2016,bib:yin2017,bib:lukin2017,Acin_2018}. At the heart of quantum technology is Quantum Optimal Control (QOC) theory \cite{bib:rabitz1988,bib:tannor1993,brif2010control,glaser2015training}.
The goal of QOC is to coherently control quantum dynamics. This is achieved by actively controlling an interaction between the system and a field (e.g. the electromagnetic field). The temporal profile of the interaction, $\omega(t)$, known as the control field, is shaped such that a given objective functional $I[\omega(t)]$ encoding the desired dynamics is minimized. A parametrization is placed on the control field and the optimal parameters are found by performing local optimization routines. 

Quantum Control Landscape (QCL) theory was developed in the early 2000s, to study the complexity involved in the search for solutions to QOC problems \cite{bib:rabitz2004}. The QCL is defined by the level hypersurfaces of the objective functional $I[\omega(t)]$. Supposing the control field is parametrized by a vector of variables $\vec{\omega}$, the Hessian matrix,

\begin{equation}
    [H(\vec{\omega})]_{i,j}=\frac{\partial{I(\vec{\omega})}}{\partial \omega_i \partial \omega_j},
\end{equation}{}

\noindent can provide fundamental topological information. It has been shown \cite{bib:shen2006,rabitz2006topology} that, for orthogonal-state-transfer control problems in  systems with finite-dimensional Hilbert space, the Hessian at a solution has an extensive null space and at most $2D-2$ non-zero eigenvalues ($D$ the dimension of the Hilbert space). Global optima constitute continuous submanifolds, level-sets of control space \cite{beltrani, moore2012exploring}. Although general results for infinite-dimensional systems are still missing, in a recent publication \cite{larocca2019exploiting}, the frequency driven Quantum Harmonic Oscillator was studied and it was proved that when targeting friction-less evolution, solutions form level-sets with at most two directions of decreasing fidelity.

The existence of continuous submanifolds of solutions has very interesting practical consequences. For example, it allows for the achievement of secondary features in the control protocols \cite{larocca2019exploiting}. The idea is the following. An initial solution to the control problem can be further optimized with respect to a new cost function, without loosing its initial fidelity, if the projection of the gradient of this auxiliary cost into the main objective solution subspace is used in a second descent procedure. Motion with this projected gradient generates fidelity-preserving trajectories with ever growing secondary yield. Let us highlight that the main ingredient for this navigation scheme are the instantaneous eigenvectors of the Hessian, particularly those associated with non-zero eigenvalues. They are crucial for the elimination of any component of the auxiliary gradient in the main objective fidelity-decreasing directions. Their computation involves the calculation of second derivatives of the objective function with respect to the control parameters. This task can become demanding in complex systems, thus limiting the applicability of the navigation method. 

In this work, we propose to bypass this difficulty by employing finite-difference (FD) approximations to compute the Hessian matrix of the cost functional for the navigation routines. FD schemes happen to be highly effective at reconstructing the Hessian eigenvectors involved in the navigation scheme. In consequence, we are able to power fidelity-preserving secondary objective gradient descents that do not require any derivatives of the cost function. As an example, we tackle Fourier Compression, an original approach that produces few-parameter smooth protocols out of irregular high-dimensional solutions. We benchmark the performance of the approximation-based scheme using two inherently different models, one finite and the other infinite-dimensional: the Landau-Zener and the Quantum Harmonic Oscillator. Both models are simple enough to allow for the exact Hessian computations, permitting to contrast the exact and approximate approaches.

The paper is organized as follows. In Section \ref{Section-Model} we introduce the two models that will be used throughout the paper. Section \ref{Section-app} is devoted to study the errors related to the approximation of the Hessian and Section \ref{Section-nav} introduces the navigation method. Section \ref{Section-fou} presents a novel navigation-powered application: the Fourier Compression. We use this secondary objective to benchmark approximate and exact navigations.  Finally, Section \ref{Section-con} holds the concluding remarks.

\section{MODELS} \label{Section-Model}

Consider the evolution of an isolated two-level quantum system, described by the Landau-Zener (LZ) Hamiltonian

\begin{equation}
\hat{H}(\omega(t)) = \frac{\Delta}{2}\hat{\sigma}_{x}+\omega(t)\hat{\sigma}_{z}
\label{ec:hami_lz}
\end{equation}

\noindent with $\sigma_{x}/\sigma_{z}$ the usual Pauli matrices, $\Delta$ the minimal energy gap, and $\omega(t)$ the control field. This model describes, for example, a qubit in a magnetic field with a fixed $x$ component and a time-dependent $z$ one. Suppose the control is initially set to $\omega=-\infty$ and the qubit is prepared on the ground state of Eq. (\ref{ec:hami_lz}), that is  $\ket{\psi(t\rightarrow-\infty)}=\ket{0}$. For a linear sweep of $\omega$ over time, $\omega_v(t)=v\:t$, an analytical formula for the asymptotic probability of finding the ground state of $H(t\rightarrow+\infty)=\ket{1}$ (that is, flipping the bit) can be derived \cite{bib:zener1932,bib:landau1932}. Since energy levels become nearly degenerate at $\omega=0$, in order to suppress the probability of populating the excited level, the level crossing has to be traversed slowly, with $v \ll v_c\propto \Delta^2$. That is, narrow crossings demand slower protocols. Recent work \cite{bib:hegerfeldt2013} have considered a finite-time version of the this problem finding that the minimum time required to traverse a crossing is given by $T_{min}=\pi/\Delta$. 

Let us introduce our objective functional, usually called infidelity,

\begin{equation}
I_{_{LZ}}[\omega(t)]=1-|\bra{1}U_T[\omega(t)]\ket{0}|^2,
\label{ilz}
\end{equation}

\noindent mapping real-valued functions $\omega(t)$ to real numbers $I_{LZ}[\omega(t)]$ measuring the departure from the target state. $U_T[\omega(t)]$ is the solution to the Schr\"odinger equation,

\begin{equation}
i\frac{d\hat{U_t}}{dt}=\hat{H}(\omega(t))\hat{U_t},
\end{equation}

\noindent evaluated at time $T$. We use piece-wise constant control fields,

\[
    \omega(t)=
    \begin{cases}
        \omega_1 & \text{if $0<t<\Delta t$}\\
        \dots \\
        \omega_M & \text{if $(M-1)\Delta t<t<T$}\\
    \end{cases}
\]

\noindent where $\Delta t=T/M$, being $T$ the duration of the protocol and $M$ the number of constant pulses. 

\begin{figure}
	\begin{center}
		\includegraphics[width=.4\textwidth]{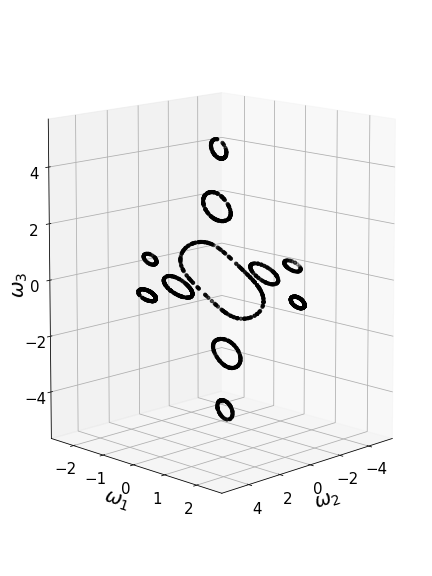}
		\caption{Solution Sets for the LZ model with $M=3$ control parameters. Each point in 3D space represents a possible protocol, with $\omega_i$ the amplitude of the $i_{th}$ pulse in the piece-wise constant sequence. We initialize thousands of random seeds and optimize with respect to the cost function of Eq. (\ref{ilz}). Those optimized fields with infidelity below $10^{-6}$ are plotted as black dots. The solutions appear to form continuous curves, one-dimensional submanifolds of parameter space.}
		\label{fig:nts3}
	\end{center}
\end{figure}

\begin{figure}
	\begin{center}
		\includegraphics[width=.4\textwidth]{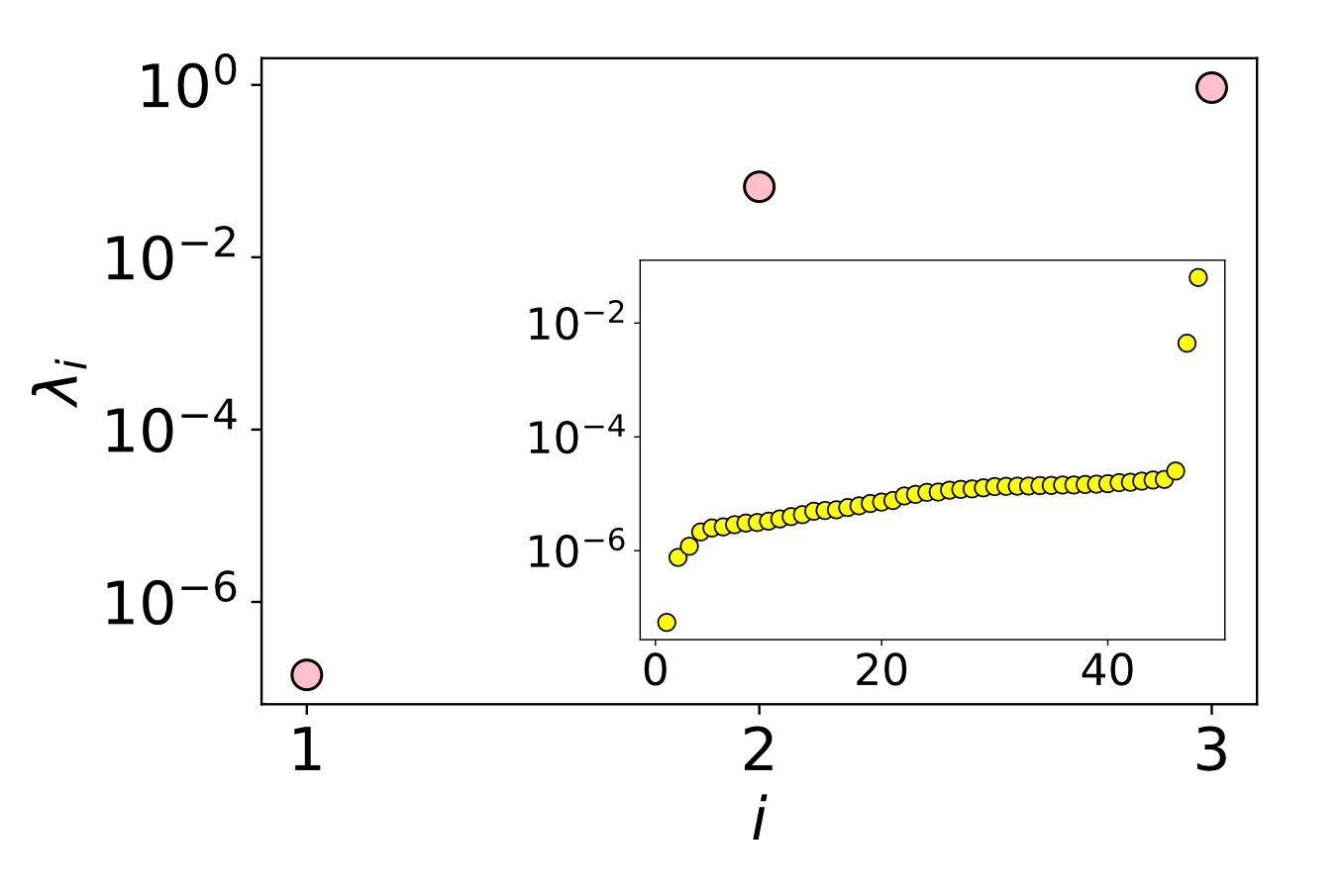}
		\caption{Eigenvalues of the Hessian of Eq. (\ref{ilz}) at a solution of M=3 (M=48 in the inset). We observe exactly two non-zero eigenvalues, corresponding to those eigendirections that depart from the solution set.}
		\label{fig:la}
	\end{center}
\end{figure}

For this model, $M_{min}=2$. If $M=2$ variables are used, multiple solutions exist but are disconnected policies in the QCL \cite{laro2018}. Instead, if we choose $M \geq 3$, continuous submanifolds of solutions arise. To better understand this situation, we consider the case $M=3$, initializing $4000$ seeds of the form $\vec{\omega}=(\omega_1,\omega_2,\omega_3)$, with $\omega_i$ randomly chosen from a uniform distribution in $[-5,5]$. The seeds are optimized with the cost function in Eq. (\ref{ilz}). In Fig. \ref{fig:nts3}, we plot those optimized fields that are globally optimal (with infidelity below a certain threshold, $I<I_{th}=10^{-6}$). A hidden structure is revealed. Solutions gather in closed loop-like formations, appearing to form continuous curves. Later, in Section \ref{Section-app}, we will show how to build a trajectory connecting these solutions. We have chosen, for the simulations, $\Delta=1$ and $T=\frac{1.4\pi}{2}$. In Fig. \ref{fig:la}, we plot the spectrum of the Hessian of one of the optimal fields that were shown in Fig. \ref{fig:nts3}. In the inset, we plot the spectrum of a $M=48$ solution. Solutions are found to have exactly two non-zero eigenvalues.

As a second model, regard a particle in a one dimensional time-dependent harmonic trap, whose evolution is described by the Quantum Harmonic Oscilator (QHO) Hamiltonian
\begin{equation}
\hat{H}(t)=\frac{\hat{p}^2}{2m}+\frac{1}{2}m\omega(t)^2\hat{x}^2
\label{ec:sho}
\end{equation}

\noindent where $\hat{x}$ and $\hat{p}$ are position and momentum operators respectively, $m$ is the mass of the particle, $\omega(t)$ is the time-dependent frequency of the trap. For simplicity, we will assume $m=\hbar=1$. Originally, the trap has frequency $\omega(0)=\omega_0$. In the context of Quantum Heat Engines \cite{salamon2009maximum,lutz,ste,stef2016minimum,Kosloff_2017}, a typical problem is to design adiabatic expansion/compression strokes. That is, finding protocols for opening/closing the trap, such that $\omega(T)=\omega_T \neq \omega_0$ and $N(T)=N(0)$, being $N$ particle number expectation value. In general, evolution with arbitrary driving protocols $\omega(t)$ causes the time-evolved Hamiltonian to no longer be diagonal in the basis of states with well defined particle number, $\hat N(0)=\hat{a}^{\dagger}(0)\hat{a}(0)$. Nevertheless, a basis diagonalizing this time-evolved Hamiltonian can be obtained through a Bogoliubov transformation \cite{larocca2019exploiting,calzetta2008nonequilibrium}

\begin{equation}
\hat{a}(T)=\alpha\hat{a}(0)+\beta\hat{a}^\dagger(0)
\label{bogo}
\end{equation}
\noindent where $\alpha$ and $\beta$ are protocol-dependent complex coefficients satisfying $|\alpha|^2-|\beta|^2=1$. Initial states with a well defined particle number (or incoherent superpositions of them) can only experience an increase in the mean particle number,

\begin{equation}
    N(T)=\left\langle \hat{a}^\dagger(T)\hat{a}(T) \right\rangle= N(0)(1+2|\beta|^2)+|\beta|^2
\end{equation}

\noindent a process called \textit{quantum friction} in the literature \cite{salamon2009maximum,lutz,ste,stef2016minimum,Kosloff_2017}. A natural measure for the departure from target friction-less evolution is given by
\begin{equation}
    I_{_{QHO}}[\omega(t)]=|\beta|^ 2
\label{I}
\end{equation}
\noindent so we will use this as the objetive functional. See Ref. \cite{larocca2019exploiting} for a detailed description on how to compute the Bogoliubov coefficient $\beta$ associated with a given driving.

\section{Approximating the Hessian} \label{Section-app}

We are interested now in studying the possibility of replacing the computation of derivatives of the cost function with approximations. Solutions to $D$-dimensional control problems have at most $M_{min}=2D-2$ directions of decreasing fidelity in the QCL \cite{beltrani}. For the LZ model, $M_{min}=2$. Ref \cite{larocca2019exploiting} demonstrated that in the QHO case, also $M_{min}=2$. The remaining $M-M_{min}$ directions are zeros of the Hessian. Consider the $\epsilon$-approximated Hessian matrix,

\begin{equation}
\begin{split}
    [H_\epsilon(\vec{\omega})]_{i,j} &\approx \frac{1}{4\epsilon^2}[I(\vec{\omega}+\epsilon(\vec{e}_i+\vec{e}_j))\\
    &-I(\vec{\omega}+\epsilon(\vec{e}_i-\vec{e}_j))\\
    &-I(\vec{\omega}-\epsilon(\vec{e}_i-\vec{e}_j))\\
    &+I(\vec{\omega}-\epsilon(\vec{e}_i+\vec{e}_i))]
\end{split}{}
\end{equation}{}



\noindent where $\vec{e}_i$ is the $i_{th}$ basis vector of parameter space and let $v_i$ and $\tilde{v}_{i}$ denote the $i_{th}$ exact and approximate Hessian eigenvectors. We will address the approximation error in each of the $M$ eigenvectors individually, defining the error

\begin{equation}
    E_i=1-v_{i}\cdot \tilde{v}_{i}
\end{equation}{}

\noindent where $\cdot$ is the standard dot product. Since degenerate subspaces cannot be associated with a unique choice of eigenvectors, there is no point in trying to measure an error associated with the degenerate null subspace. Moreover, the projection operation only requires those eigenvectors associated with non-zero eigenvalues, such that any component of the motion out of the null subspace can be neglected.

\begin{figure}
	\begin{center}
		\includegraphics[width=.4\textwidth]{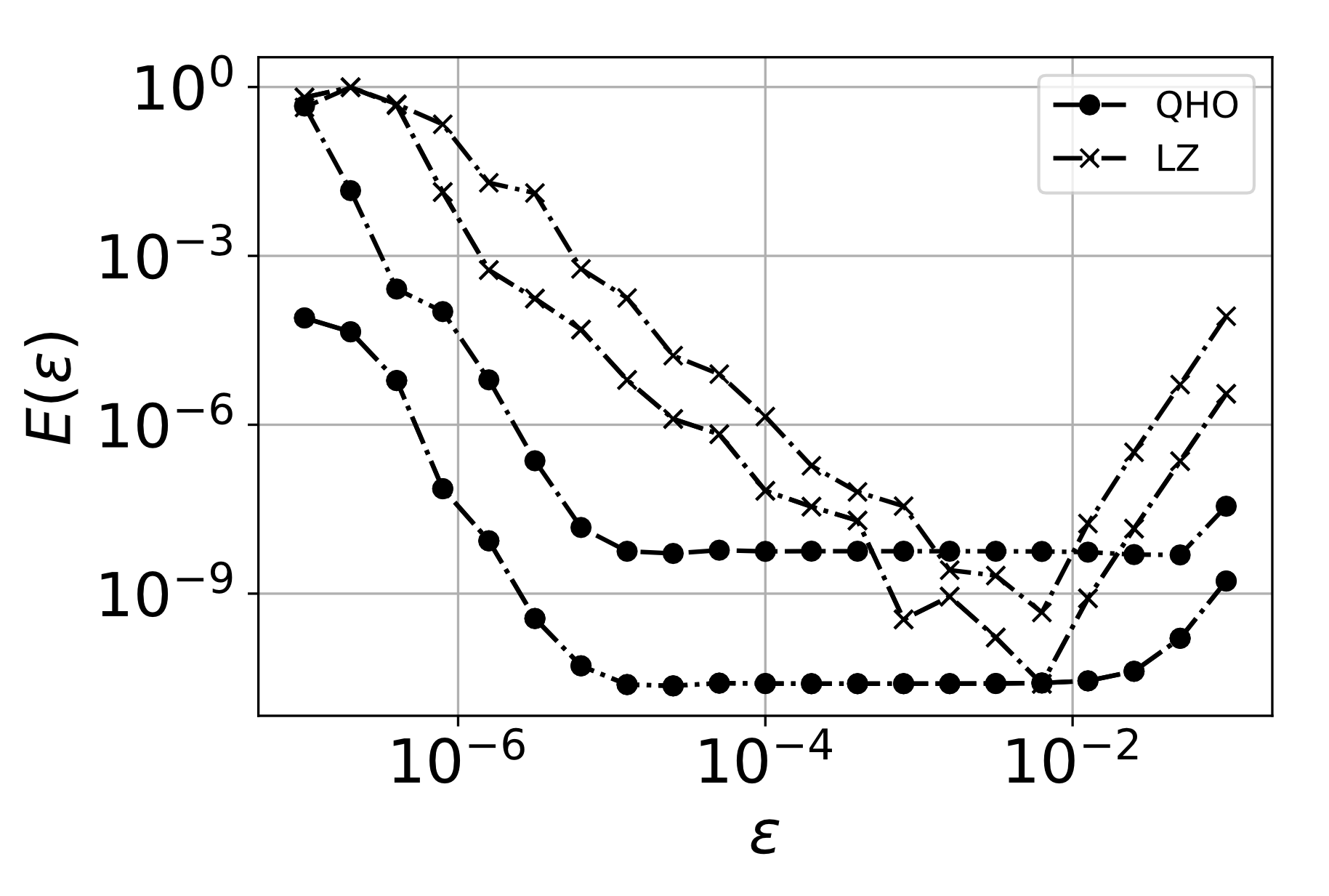}
		\caption{Overlap error $E(\epsilon)$ between exact and approximate Hessian eigenvectors, as a function of the step in the FD computation, $\epsilon$, for the QHO model (dots) and the LZ model (crosses). Only those eigenvectors associated with non-zero eigenvalues are presented.}
		\label{dip}
	\end{center}
\end{figure}

We initialize one hundred $M=6$ solutions and for each of them we build two curves, one for each non-zero eigenvector error $E_i$ as a function of $\epsilon$, the step in the FD computation. In Fig. \ref{dip} (a) and (b) we present these errors for the LZ model and the QHO respectively. Although we choose to plot the curves corresponding to only one representative solution for each model, the observed behaviour was similar for the remaining controls analyzed. In both cases, excellent approximations are found, with errors below $10^{-8}$. Here and onwards, total evolution time is set to $T=1.8$.





\section{FINITE-DIFFERENCE NAVIGATION}\label{Section-nav}
In the previous section we have shown that the Hessian eigenvectors, needed by the navigation method, can be accurately approximated using finite-differences. In this Section, we put in practice such approximation-based routine to navigate through solutions. 

\begin{figure}
	\begin{center}
		\includegraphics[width=.4\textwidth]{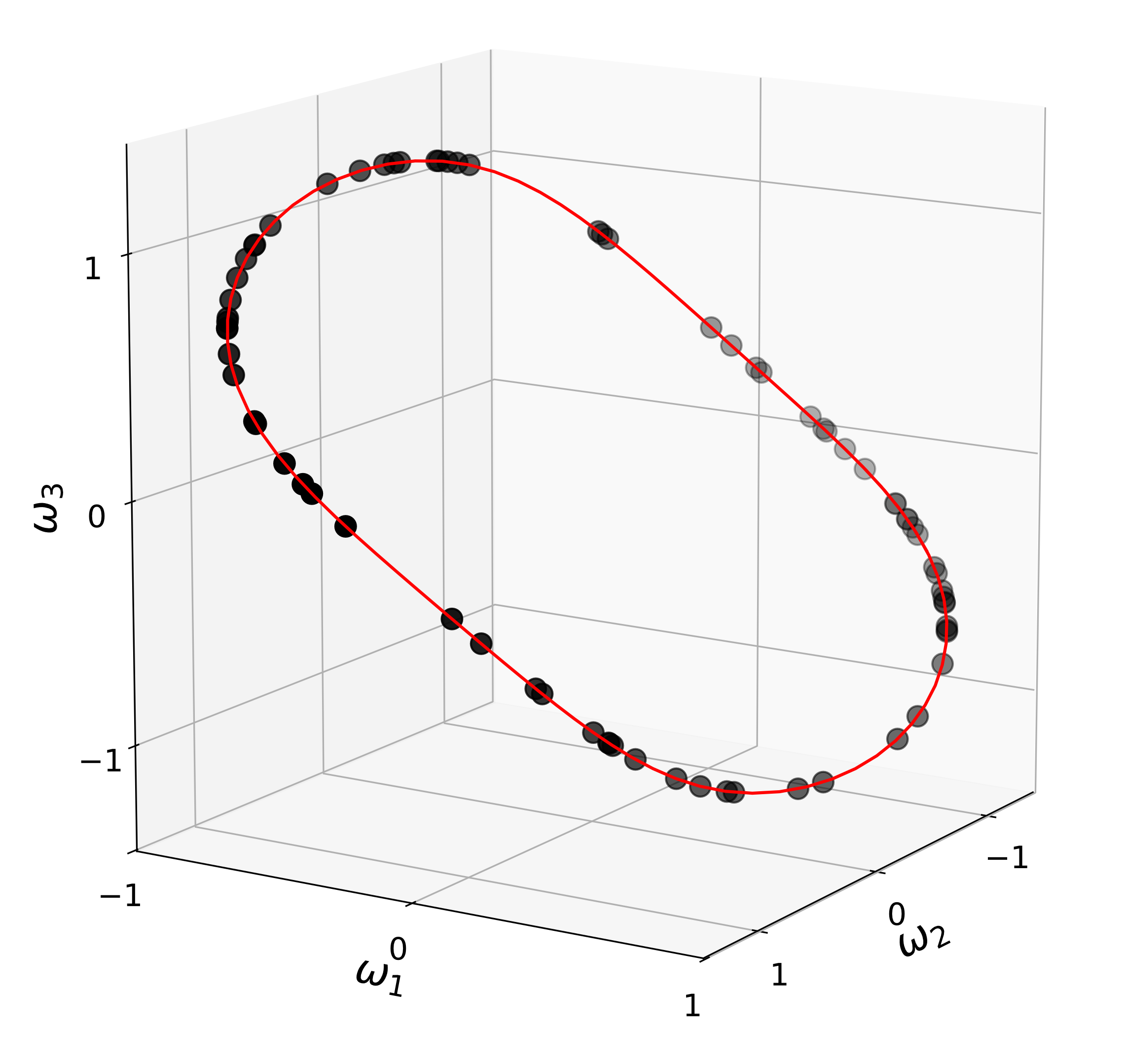}
		\caption{Following the Null Hessian Eigenvector. Black dots depict solutions found optimizing thousands of initial random seeds (see Fig. (\ref{fig:nts3})). Starting from one of these solutions, we follow the FD-generated Hessian eigenvector associated with the null eigenvalue generating a fidelity preserving trajectory (red curve) linking all of the scattered solutions.}
		\label{fig:drive}
	\end{center}
\end{figure}

Consider a trajectory $\vec{\omega}(\zeta)$, solution to an initial value problem,

\begin{equation}
    \frac{d\vec{\omega}(\zeta)}{d\zeta}=f(\vec{\omega}(\zeta))
\end{equation}

\noindent with $\vec{\omega}(0)=\vec{\omega}_0$. If we choose $f(\vec{\omega}(\zeta))$ to be $P\vec{a}$, with $\vec{a}$ an arbitrary vector and $P$ the following projection operation 

\begin{equation}
    P\vec{a}=\vec{a}-\sum_{i/\lambda_i\neq0} \vec{v}_ic_i
\label{proj}
\end{equation}{}


\noindent where the sum runs through the non-zero Hessian eigenvectors and $c_i=\vec{a}\cdot\vec{v_i}$, the arising trajectory is constrained to solution space. We use a fourth-order Runge-Kutta integration routine to numerically approximate these trajectories. In Fig. \ref{fig:drive} we demonstrate how the approximate Hessian can be used to drive through one of the solution level-sets of Fig. \ref{fig:nts3}. Starting from an optimal field, a perfect fidelity trajectory (red line), is built by following, at each iteration, the instantaneous eigenvector associated with the null Hessian eigenvalue (see Fig. \ref{fig:nts3}). In the simulations, $\epsilon=10^{-2}$ and $h=0.01$.

\begin{figure}
	\begin{center}
		\includegraphics[width=.45\textwidth]{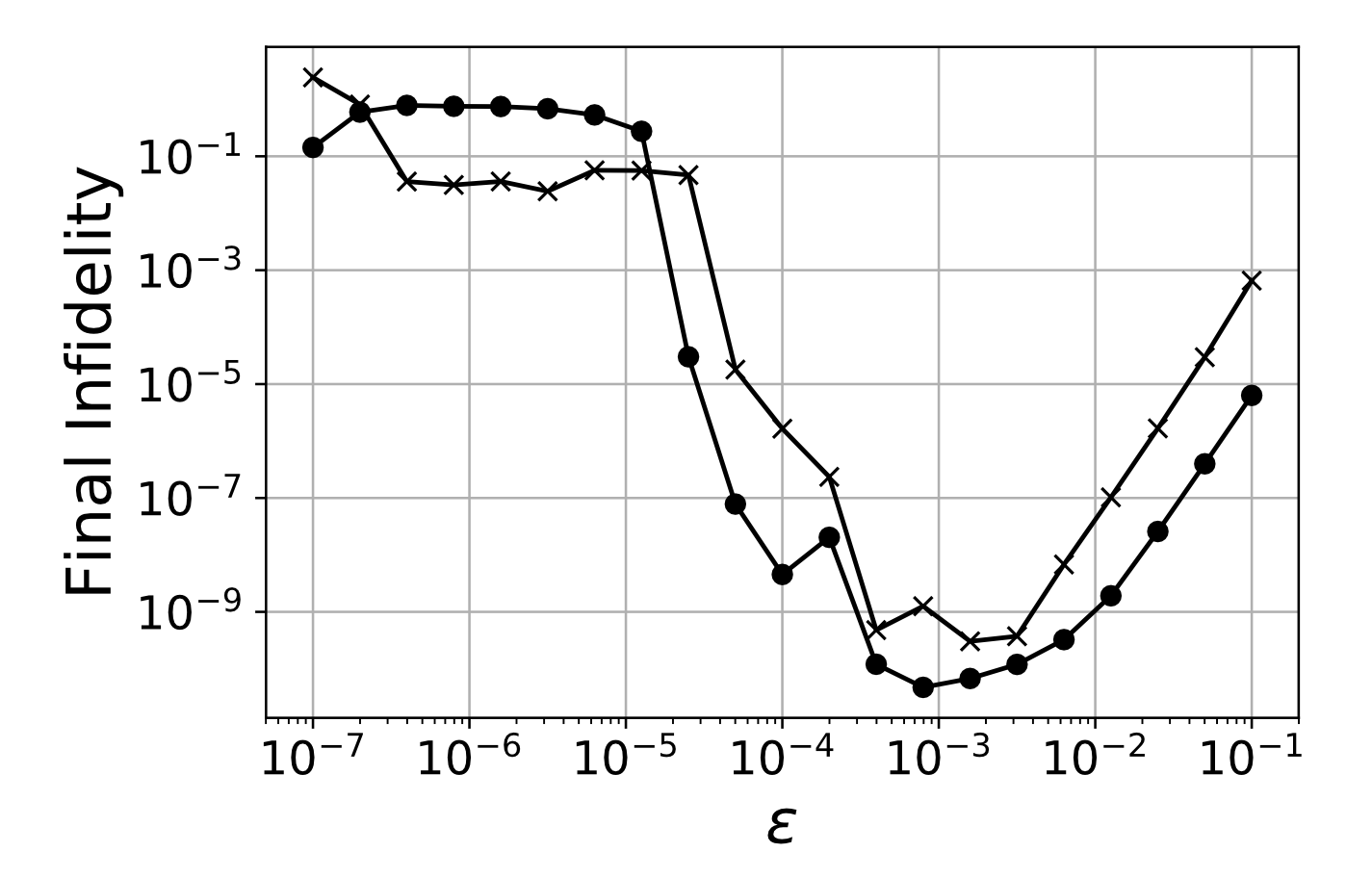}
		\caption{Calibration Procedure. A solution to the QHO problem with $M=6$ parameters is initialized. A direction in control space is chosen at random and a navigation sequence following its projection onto solution space is started. After $1000$ iterations with a step of $h=0.1$, the final infidelity is recorded. The solution is initialized again and a navigation following the same randomly selected direction, this time with a new step in the FD scheme is initiated. The final infidelity is plotted as a function of $\epsilon$ (black dots), allowing for the selection of optimal $\epsilon$ without having to compute the exact eigenvectors (see Fig. \ref{dip}). A second random direction (black crosses) yields similar results.}
		\label{cali}
	\end{center}
\end{figure}

Although Fig. \ref{dip} shows that non-degenerate eigenvectors can be faithfully built with FD schemes for a wide range of values for parameter $\epsilon$, a natural question arises: is there a way of choosing a proper $\epsilon$ without knowledge of the exact Hessian? Consider the following $calibration$ procedure. A solution is initialized and a random direction in parameter space is chosen. A trajectory following the projection of this random direction into solution space is generated and the final infidelity is recorded. Ideally, if the step in the trajectory is sufficiently small and the eigenvectors involved in the projection operation are faithfully approximated, the infidelity along the trajectory will remain optimal. Restarting the initial solution and building new trajectories for different values of $\epsilon$, we plot the final infidelity as a function of this parameter (see Fig. \ref{cali}). Two different $M=6$ solutions to the QHO problem were tested (dots and crosses), running $1000$  iterations. The fundamental behaviour in Fig. \ref{dip} is recovered, this time without any computation of the exact Hessian. 

\begin{figure}
	\begin{center}
		\includegraphics[width=.45\textwidth]{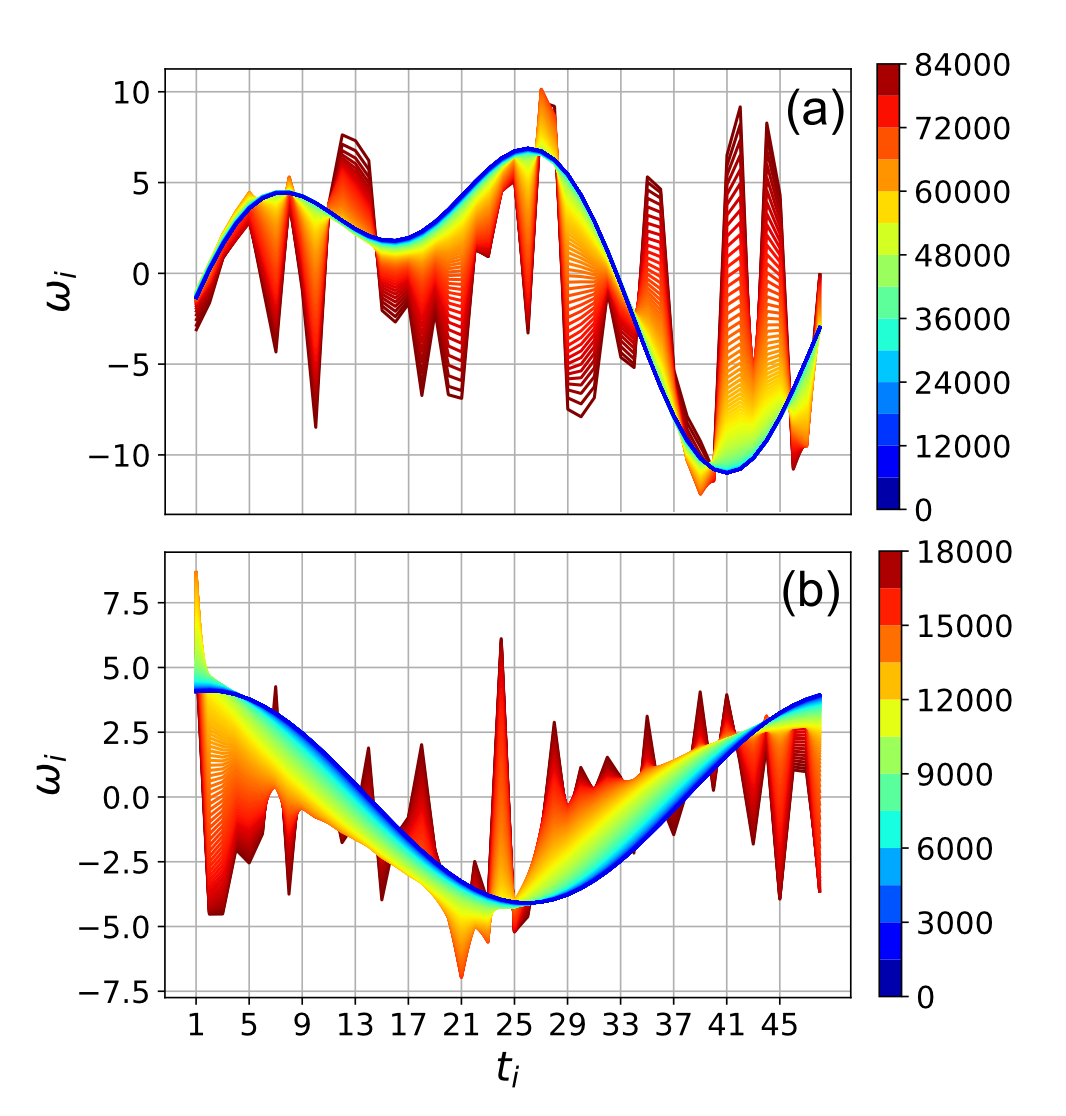}
		\caption{Fourier Protocol Compression in the QHO. Each point in the graph depicts the value of the $i_{th}$ component of a given protocol, corresponding to the time interval $\Delta t_i$. That is, the curves represent distinct protocols, which were coloured relative to their secondary objective cost value of Eq. (\ref{foucost}). All of them are optimal with respect to the main objective in Eq. (\ref{I}). Starting with different random solutions, we present Fourier compression trajectories with \textbf{(a)} $\vec{p}=(1,2)$ , and \textbf{(b)}  $\vec{p}=(1)$.}
		\label{fig:QHO}
	\end{center}
\end{figure}

\begin{figure}
	\begin{center}
		\includegraphics[width=.45\textwidth]{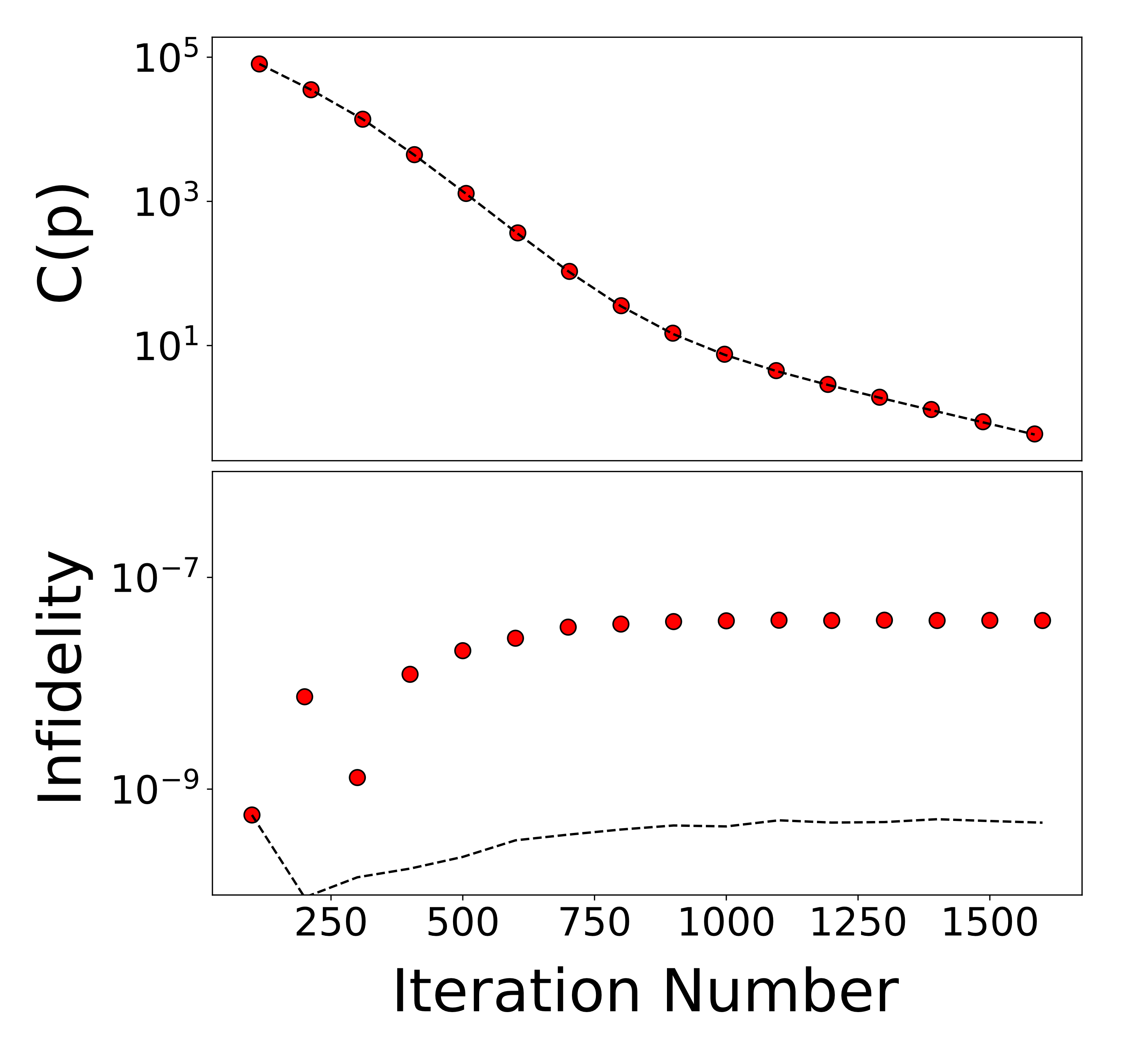}
		\caption{Exact and Approximate Trajectories Compared.  (a) Fourier Cost and (b) infidelity evolution in the exact (dotted black lines) and approximate (scattered red circles) trajectories of Fig. \ref{fig:QHO}.(a). Cost trajectories are indistinguishable while the infidelity remains below $10^{-7}$ in both cases.}
		\label{fig:fouafou}
	\end{center}
\end{figure}

\section{FOURIER COMPRESSION}\label{Section-fou}
In a previous work \cite{larocca2019exploiting}, we have demonstrated how to exploit the navigation procedure to smooth or compress solutions obtained from raw optimization. $Smoothing$ procedures were shown to be effective at producing regular control fields. Alternatively, protocols described by only a few parameters were obtained putting forward $compression$ procedures. Since compression was performed in the original, piece-wise constant parametrization, the two approaches were incompatible. Let us show how these two secondary features can be simultaneously tackled by extending the compression idea to the frequency realm. First, let us introduce the Discrete Fourier Transform (DFT). The DFT maps an M-dimensional vector $x$ into another M-dimensional complex vector $X$ with its $k_{th}$ component given by

\begin{equation}
    X_k=\sum_{n=0}^{M-1} x_n e^{-\frac{i2\pi k n}{M}}
\end{equation}{}

\noindent That is, the $k_{th}$ component of X encodes the projection of x into the complex exponential with frequency $k$. Now suppose we want our protocol to have only a finite number of frequencies

\[
    \begin{cases}
        |X_k|^2 \neq 0 & \text{if $k \in \vec{p}$} \\
        |X_k|^2= 0 & \text{if $k \not \in \vec {p}$}\\
    \end{cases}
\]

\noindent for some arbitrary $\vec{p}$, for example $\vec{p}=(1,2)$. We define the auxiliary cost of having exactly the frequencies in $\vec{p}$ as

\begin{equation}
    C(\vec{p})=\sum_{k\not \in \vec{p}}^{M-1} |X_k|^2
\label{foucost}
\end{equation}{}

Beginning with random $M=48$ solutions to the QHO control problem, in Fig. \ref{fig:QHO} we follow the gradient of Eq. (\ref{foucost}) projected onto the solution subspace (as explained in Eq. \ref{proj}) for (a) $\vec{p}=(1,2)$ and (b) $\vec{p}=(1)$. To further validate the method, in Fig. \ref{fig:fouafou} we show the evolution of the cost and the infidelity along the approximate trajectory (red circles) and compare it with the exact secondary descent (dotted lines). Albeit greater, the infidelity in the approximate trajectory remains practically zero. Notably, the secondary cost trajectories are identical. Of course, there is a run-time advantage in using the exact Hessian, but it is not a categorical difference. In average, we found the approximate routine to perform $2.7$ times slower. Similar trajectories for the FD-based Fourier Compression in the LZ case are presented in Fig. \ref{fig:lz}. 

Finally, a run-time benchmarking study is presented in Fig. \ref{rt}. Defining the run-time spent in finding the Hessian eigenvectors with the approximate and the exact approaches, $\tau_{app}$ and $\tau_{ext}$ respectively, we initialize $100$ random controls for different values of $M$ and plot the mean value of the quotient $\eta=\frac{\tau_{app}}{\tau_{ext}}$ for both the LZ (red squares) and QHO model (black circles). Exact and approximate run-times are found to be of the same order of magnitude, in both cases.

\begin{figure}
	\begin{center}
		\includegraphics[width=.45\textwidth]{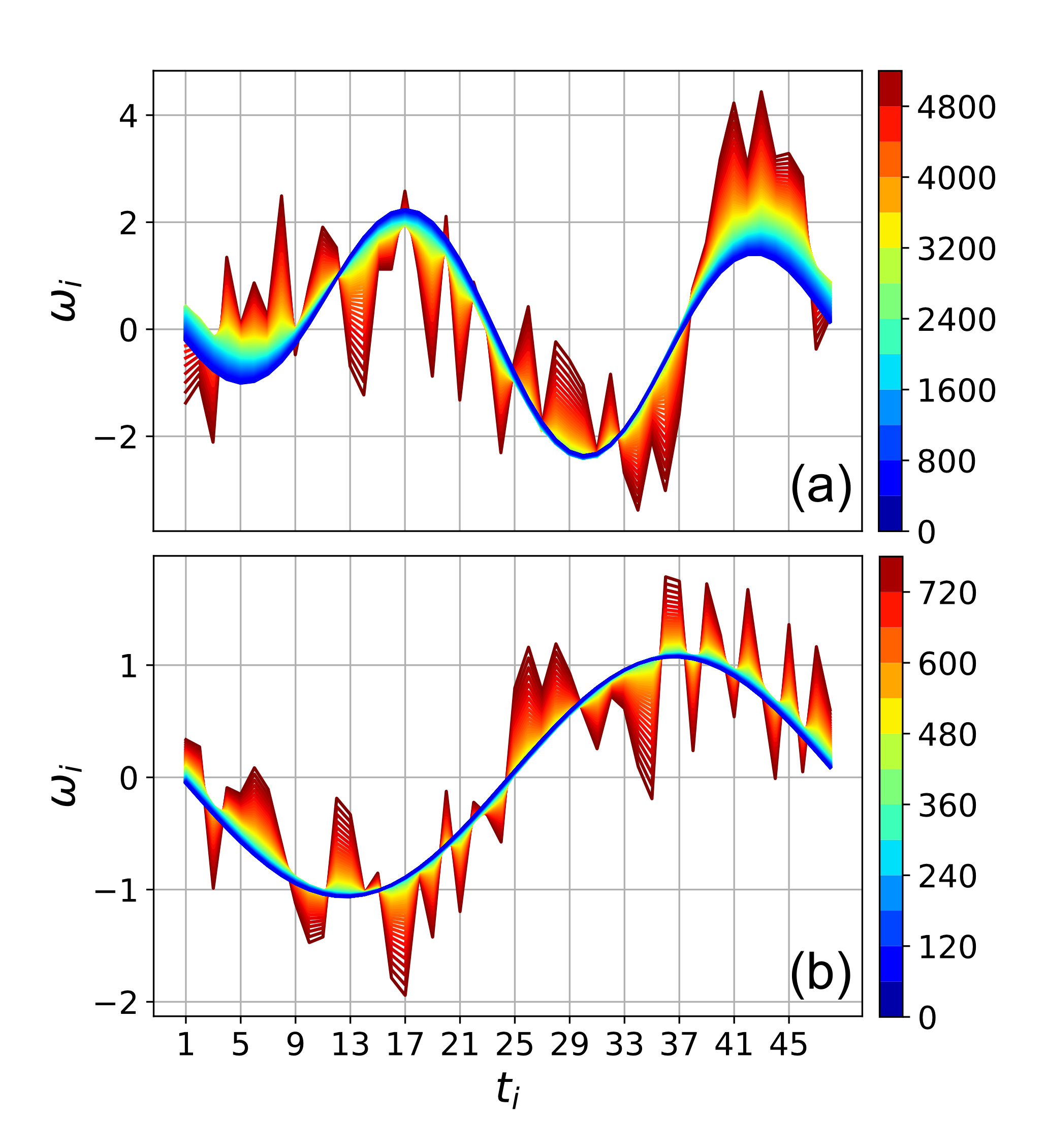}
		\caption{Fourier Protocol Compression in the LZ model. Each point in the graph depicts the value of the $i_{th}$ component of a given protocol, corresponding to the time interval $\Delta t_i$. The curves represent distinct protocols, which were coloured relative to their secondary objective cost value of Eq. (\ref{foucost}). All of them are optimal with respect to the main objective in Eq. (\ref{ilz}). Starting with different random solutions, we present Fourier compression trajectories with \textbf{(a)} $\vec{p}=(1,2)$ , and \textbf{(b)}  $\vec{p}=(1)$.}
		\label{rt}
	\end{center}
\end{figure}

\begin{figure}
	\begin{center}
	\vspace{1cm}
		\includegraphics[width=.35\textwidth]{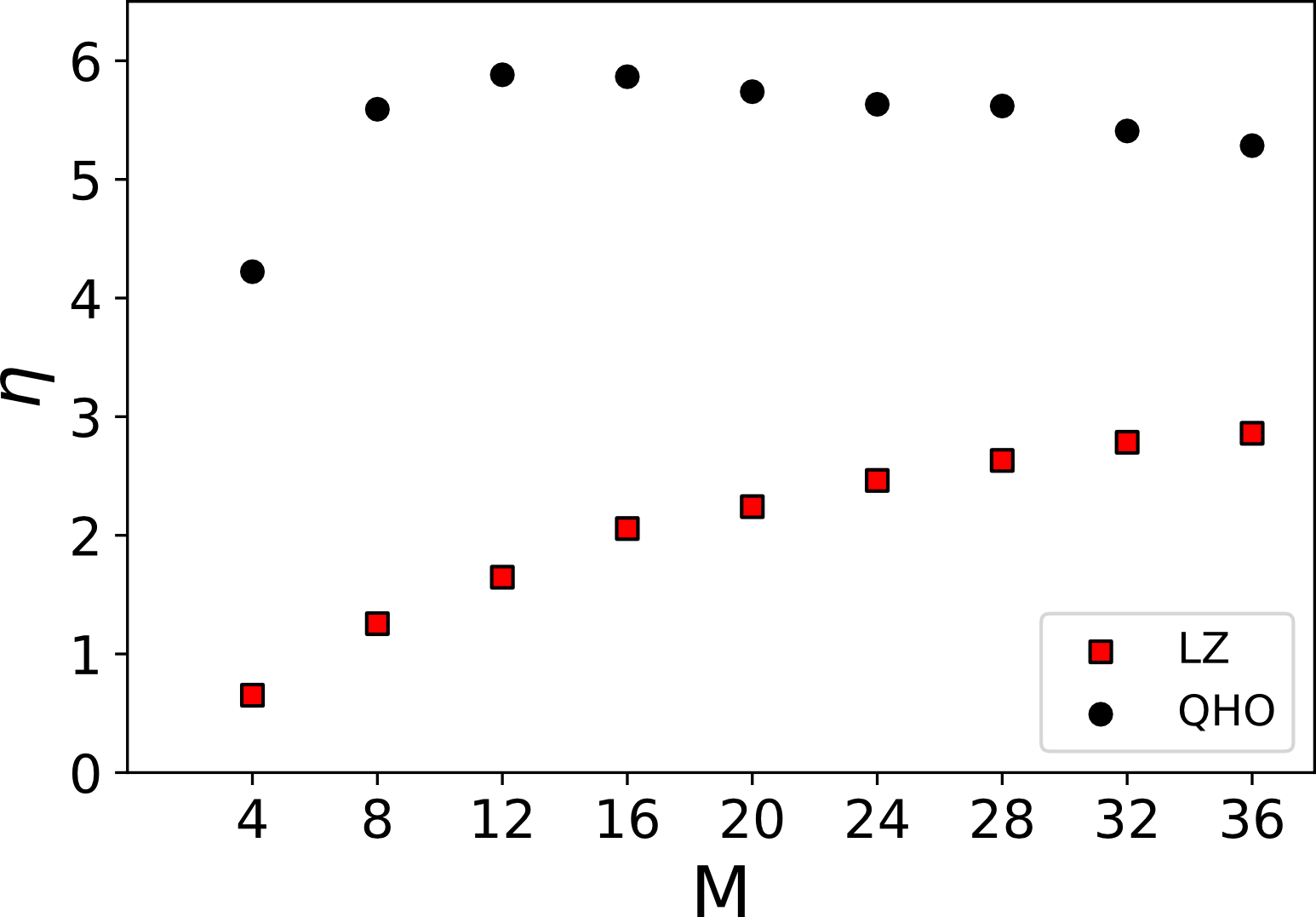}
		\caption{Run-time Analysis. $100$ random fields are initialized for different values of $M$, the Hessian eigenvectors are computed and the run-time is recorded. The mean quotient between approximate and exact run-times,$\eta=\frac{\tau_{app}}{\tau_{ext}}$, is plotted for the LZ model (red squares) and the QHO model (black circles). In the two scenarios, we find the same order of magnitude for exact and approximate run-times.}
		\label{rt}
	\end{center}
\end{figure}

\section{FINAL REMARKS}\label{Section-con}

Solutions to quantum control problems form continuously varying level-sets in the QCL. Recent work evidenced that this property may have major practical consequences \cite{larocca2019exploiting}. Navigation methods were shown to provide a novel, straight-forward way of producing secondary features in the control protocols (e.g. smoothness or compression). Unfortunately, the method required the computation of the second derivatives of the cost functional, a task that may become prohibitively expensive in complex systems.

In this work, we analyzed the possibility of a derivative-free approach to the navigation method, utilizing finite-difference approximations instead of the exact Hessian computations. First, we showed that the eigenvectors of the Hessian can be accurately constructed with FD approximations. Having characterized the error in the eigenvectors, we explicitly used the FD eigenvectors to generate optimal trajectories in parameter space.

In particular, we tested the performance of the FD scheme at executing secondary objective optimizations. Defining an auxiliary cost penalizing the Fourier components of the control protocol, initially irregular and high-dimensional solutions were evolved into smooth controls described by only a few parameters. Fourier Compression constitutes a novel approach to smoothness and compression in quantum control.


Although originally developed to explore the complexity of the search for controls, the study QCL's is proving to be more fruitful than expected. The results presented in this work pave-the-way for the design of universal solution subspace navigation methods, extending their application beyond analytically solvable models. 

\begin{acknowledgements}
The authors acknowledge financial support from \mbox{ANCyPT} (Grant no. PICT-2016-1056), \mbox{CONICET} (Grants No PIP 11220170100817CO and PIP 11220150100493CO), and UBACyT (Grants No 20020130100406BA and 20020170100234BA).

\end{acknowledgements}

\bibliography{laro2020.bib}	
\end{document}